\documentstyle[preprint,aps]{revtex}  

\begin{document}

\draft

\title{The effective potential of gauged NJL model in
magnetic field}

\author{I. Brevik }
\address{Dept. of Applied Mechanics and 
 Norwegian Institute of Technology\\ 
Univer.  of Trondheim, Trondheim, Norway }

\author{D.M. Gitman\thanks{e-mail: 
gitman@fma.if.usp.br}}

\address{Instituto de F\'{\i}sica, Universidade de S\~ao Paulo \\ 
Caixa Postal 66318, 05315-970-S\~ao Paulo, S.P., Brasil
}
\author{S.D. Odintsov\thanks{On leave from  Tomsk Pedagogical
University, 634041 Tomsk, Russia; present  
e-mail: 
odintsov@fma.if.usp.br}}
\address{Instituto de F\'{\i}sica, Universidade de S\~ao Paulo
 \\ Caixa Postal 66318, 05389-970-S\~ao Paulo, S.P., Brasil;\\
Universidad del Valle, Cali, Colombia}

\date{\today}

\maketitle

\begin{abstract} 
The formalism, which permits to study the phase structure of
gauged NJL-model for arbitrary external fields is developed. The
effective potential in the gauged NJL model in the weak magnetic field
is found. It is shown that in fixed gauge coupling case the weak
magnetic field doesn't influence chiral symmetry breaking
condition. The analogy with the situation near black hole is briefly
mentioned.
\end {abstract} 
  
\newpage

\section {Introduction}

For a long time it is well known that QED may have a non-perturbative
strong-coupling phase \cite{1} where the chiral symmetry is
broken. The manisfestation of this phase maybe found via analysis
\cite{2} of the multiple correlated and narrow-peak structures  in
electron and positron spectra \cite{3}. Schwinger-Dyson (SD) equations
approach is usual tool to study such a strong-coupling phase.

The gauged Nambu-Jona-Lasinio (NJL) model represents even more
interesting theory where such a non-perturbative strong-coupling
chiral symmetry broken phase maybe naturally realized. Moreover,
gauged NJL-model maybe used to describe the Standard Model. It may
also play the role of low-energy effective theory for QCD. It is very
interesting to understand what external conditions may influence the
chiral symmetry breaking in such theory. Among such an external
conditions the external magnetic field could be very interesting one,
as it can be realized in the laboratory experiments. 

There has been recently some interest in the study of (non-gauged) NJL
model in an external magnetic field \cite{4}. It has been shown that
strong magnetic field increases the value of dynamically generated
fermionic mass, and in this way, it supports the phase with chiral
symmetry breaking. Such a stydy is closely related with the
investigation of effective potential in an external magnetic fields
\cite{20} where the possibility of symmetry breaking due to external
magnetic field has been shown.

In the present work we discuss the effective potential in the gauged
NJL model in an external magnetic field. The explicit evaluation of
the effective potential is given in the situation  when magnetic field
is weak. It is shown that weak magnetic field doesn't influence chiral
symmetry breaking.

\section{Gauge Higgs-Yukawa model in  external magnetic field}

We will start from the $SU(N_c)$ gauge theory with scalars and
spinors:
\begin{eqnarray}\label{1} 
L_m =&-&{\frac{1}{4}}~ G^a_{\mu\nu} G^{a\mu \nu} ~+~ {\frac{1}{2}}~ 
g^{\mu\nu}\partial_{\mu}\sigma \partial_{\nu}\sigma 
 ~-~\frac{1}{2}~ m^2\sigma^2~-~{\frac{\lambda}{4}}~\sigma^4 \nonumber\\ 
&+&~\sum^{N_f}_{i=1}~ \bar{\psi}_i ~i\hat{D}~ \psi_i ~-~  
\sum^{n_f}_{i=1}~y\sigma \bar{\psi}_i\psi_i~.~ 
\end{eqnarray} 
\noindent
where $\sigma$ is a single scalar, $y$ is Yukawa coupling and $N_f$
fermions belong to the representation $R$ of group $SU(N_c)$.
Moreover, $n_f << N_f$, i.e. only $n_f$ fermions have large Yukawa
couplings.

We will briefly describe now the modified $1/N$ approximation \cite{5}
to study the theory (\ref{1}) at high energies. 

\noindent i) As usualy in $1/N$ - approximation the gauge coupling constant is
supposed to be small 
\[
\frac{g^2 N_c}{4 \pi} << 1\;.
\]

\noindent
ii) The number of fermions should be comparable with $N_c$
\[
N_f \sim N_c \;.
\]
\noindent
iii) We are working in the leading order of $1/N$ expansion,
i.e. scalar loop contribuitions should be negligible
\[
\left| \frac{\lambda}{y^2}\right|\leq N_c \;.
\]

Working in frames of such modified $1/N$ approsimation \cite{5}, one
can write the standard one-loop RG equations for the coupling
constants,
\begin{eqnarray}\label{2} 
&{}&{\frac{dg(t)}{dt}}=-~{\frac{b}{(4\pi)^2}}~g^3(t)~, 
\nonumber\\ 
&{ }&{\frac{dy(t)}{dt}}= 
        {\frac{y(t)}{(4\pi)^2}}~[a~y^2(t)-c~g^2(t)]~, 
\nonumber\\ 
&{ }&{\frac{d\lambda (t)}{dt}} =   
        {\frac{u~y^2(t)}{(4\pi)^2}}~[\lambda (t)-y^2 (t)]~.
\end{eqnarray} 
  
\noindent Here $ b = (11 N_c-4T(R)N_f)/3, c = 6~C_2(R), a = u/4 =
2~n_fN_c$. For example, in fundamental representation $T(R) = \frac{1}{2},\; C_2(R)
= \frac{N^2_c - 1}{2N_c}$. As usually, renormalization group
parameter $t=\ln\frac{\mu}{\mu_0}$.

Let us consider now the situation when the theory (\ref{1}) interacts
with the external magnetic field \cite{8}. It means that spinor
derivative in (1) contains the external magnetic field
connection. Then, one can calculate the effective potential $ V(\sigma)$
for the constant scalar field $\sigma$. Working in the modified $1/N_c$
- approximation and treating the external magnetic field exactly we
get:
\begin{equation}\label{3} 
V={\frac{1}{2}} m^2 \sigma^2 + {\frac{\lambda}{4}} \sigma^4 + 
 {\frac{a}{(4\pi)^2}}\int_{1/\Lambda^2}^{\infty}\frac{ds}{s^2}e^{-sM^2_F}\,
{\rm coth}(eHs)\,eH\;, 
\end{equation} 
\noindent
where  $ e $ is electric charge, $H $ is magnetic field $M_F = y\sigma $
plays the role of fermionic mass, $ \Lambda $ is cut-off. Note that
effective potential (\ref{3}) is written in regularized form but not in
renormalized form.

In order to write the effective potential in renormalized form we will
make the expansion in powers of $ H$ (derivative expansion) following
the technique of ref \cite{20} (taking account also of renormalization
 conditions)
\begin{eqnarray}\label{4} 
V&=&{\frac{1}{2}} m^2 \sigma^2 + {\frac{\lambda}{4}} \sigma^4 
- {\frac{a M^4_F}{2(4\pi)^2}} \left[\ln  
{\frac{M^2_F}{\mu^2}}-{\frac{3}{2}}\right]\nonumber\\ 
&{}& -~{\frac{a e^2 H^2}{24(\pi)^2}} \left[\ln 
{\frac{M^2_F}{\mu^2}}\right]\;, 
\end{eqnarray} 
\noindent
where we suppose that $M^2_F > eH$. Here, third term gives the
standard Coleman-Weinberg one loop potential \cite{7}, and last term
in (\ref{4}) is Schwinger's effective potential.

One can also study the case when the external magnetic field being
weak is bigger than scalar: $ eH > M^2_F,\; eH < 1$. Then, one can get
(compare with \cite{10})
\begin{eqnarray}\label{5} 
V&=&{\frac{1}{2}} m^2 \sigma^2 + {\frac{\lambda}{4}} \sigma^4 
- {\frac{a M^4_F}{2(4\pi)^2}} \left[\ln  
{\frac{eH}{\mu^2}}-{\frac{3}{2}}\right]\nonumber\\ 
&{}& -~{\frac{a e^2 H^2}{24(\pi)^2}} \left[\ln 
{\frac{eH}{\mu^2}}\right]\;. 
\end{eqnarray} 
Generally speaking, one should also add to effective potential
(\ref{4},\ref{5}) the classical potential of magnetic field
\[
V_{cl}^H=\frac{1}{4}F^2_{\mu\nu}=\frac{1}{2}H^2\;.
\]
Similarly, one can consider other regimes for external magnetic field.

The analysis of RG equations for gauge, Yukawa and scalar coupling
constants has been given in \cite{5}. It has been shown that theory
maybe non-trivial and stable in above approximation (gauge coupling
and Yukawa coupling constants will be assymptotically free). On the
same time, there will be non-trivial solution for scalar coupling
constants \cite{5} only when $c>b$. Note that if one considers the
coupling constants space then the solution of \cite{5} lie on the
line between the Gaussian fixed point and the fixed point of
\cite{11}.

\section{Gauged NJL model}
Let us turn now to the $SU(N_c)$ gauged NJL-model with four-fermion
coupling constant G. The corresponding Lagrangian maybe written as
follows:
\begin{equation}\label{6} 
L=-{\frac{1}{4}} G^2_{\mu \nu} +i {\sum\limits^{N_f}_{i=1}} 
\bar{\Psi}_i  \hat{\cal D} 
\Psi_i + G ~{\sum\limits^{n_f}_{i=1}} (\bar{\Psi}_i \Psi_i)^2~. 
\end{equation} 
As usually the way to study such a model consists in the rewriting of
Lagrangian (\ref{6}) in terms of equivalent theory with an auxiliary
scalar field $\sigma$ (fermionic condensate). The latter theory should
be identified with the Higgs-Yukawa model.

As it has been shown in \cite{12} RG approach maybe applied in
order to construct such an identification. In this approach one can
put a set of boundary conditions for the effective couplings of the
gauge Higgs-Yukawa model at $t_\Lambda = \ln \frac{\Lambda}{\mu_0}$
 (where $\Lambda$ is $UV$ cut-off) in order to prove the equalence of
gauge Higgs-Yukawa model and gauged NJL-model. More exactly, one
proves \cite{5} that sequence of gauge Higgs-Yukawa theories parametrized by
$\Lambda$ is  equivalent to the corresponding sequence of gauged NJL
models. Moreover, this equivalence is kept even at $\Lambda \rightarrow
\infty$. Because gauge Higgs-Yukawa theory is renormalizable, the
gauged NJL model maybe also called renormalizable in this sense
\cite{5}.

We will consider below only fixed gauge coupling case, $b \rightarrow +
0$. That case reminds us so-called asymptotically finite theories
\cite{13}. The general case for arbitrary $b$ can be also
done. However the corresponding expressions can not be written in
explicit form so they are not vety instructive.

The solution of first RG Eg. (\ref{2}) is the same in gauge
Higgs-Yukawa theory, or in gauged NJL model:
\begin{equation}\label{7} 
\eta (t)\equiv {\frac{g^2(t)}{g^2_0}}\equiv {\frac{\alpha 
(t)}{\alpha_0}}= 
(1+{\frac{b~ \alpha_0}{2\pi}}t)^{-1}~. 
\end{equation} 
For fixed gauge coupling limit $b\rightarrow +0$ we get 
\begin{equation}\label{8}
\eta^{-c/b}(t)_{b\rightarrow 0}{\rightarrow}\exp
(\frac{\alpha}{\alpha}_c t)=(\frac{\mu}{\mu_0})^{\alpha/\alpha_c}\;,
\end{equation}
where $\alpha\equiv \alpha_0$ is the initial value for gauge coupling
and 
\begin{equation}\label{9}
\alpha^{-1}_c \equiv \frac{c}{2 \pi} = \frac{3C_2(R)}{\pi}\;.
\end{equation}

For scalar and Yukawa couplings we should solve RG eqs. (\ref{2}) in
general case. Then, the compositeness conditions \cite{12} should be
used in order to obtain running Yukawa and scalar couplings in gauged
NJL model for $b \rightarrow +0$. The result looks as following
\cite{5}
\begin{eqnarray}\label{10} 
&{}& y^2 (t) ={\frac{(4\pi)^2}{2a}}~  \frac{\alpha} 
{\alpha_c} \left[ 
1-\left({\frac{\mu} 
{\Lambda}}\right)^{\frac{\alpha} 
{\alpha_c} } \right]^{-1} 
\equiv  
y_\Lambda^2 (t)\;,\nonumber\\ 
&{}& {\frac{\lambda (t)}{y^4(t)}} = 
{\frac{2a}{(4\pi)^2}}~~{\frac{\alpha_c} 
{\alpha}} 
\left[ 1-  \left({\frac{\mu} 
{\Lambda}}\right)^{\frac{2\alpha} 
{\alpha_c} }  \right] \equiv {\frac{\lambda_\Lambda 
(t)}{y^4_\Lambda(t)}}~, 
\end{eqnarray} 
where $t<t_\Lambda$ and condition $c>b$ is automatically satisfied
because $c$ is a positive constant. Note that in the limit $\Lambda
\rightarrow \infty$, running coupling constants (\ref{10}) tend to the
fixed points of ref. \cite{11}.

In addition, we should use also a compositeness condition for the
mass. It leads to the following running mass \cite{5}:
\begin{equation}\label{11} 
m^2 (t) = {\frac{2a}{(4\pi)^2}} \left( 
{\frac{\Lambda^2}{\mu^2}}\right)^W 
y^2_\Lambda (t)~\mu^2 \left[{\frac{1}{g_4(\Lambda)}} - 
{\frac{1}{W}}\right]~, 
\end{equation} 
\noindent
where $W \equiv 1 - \frac{\alpha}{2 \alpha_c}$ and
$g_4(\Lambda)$ is dimensionless constant defined by
\begin{eqnarray}\label{12}
G \equiv  \left( \frac{(4 \pi)^2}{a} \frac{g
_4(\Lambda)}{\Lambda^2}\right)\;.
\end{eqnarray}
Note that in case of an external gravitational field one has extra
compositeness conditions, for example, for scalar-gravitational
coupling \cite{10} (for non-gauged NLJ-model, see also \cite{14}).

To conclude, we obtained the description of the gauged NJL model (in
fixed gauge coupling case) via correspondent running coupling
constants. These running couplings were obtained using the equivalence
with the gauge Higgs-Yukawa model.

\section{RG improved effective potential and dynamical chiral
symmetry breaking}

We will be interested here in the calculation of  the effective
potential for gauged NJL-model in an external magnetic field. The
analysis of such an effective potential gives the answer to the
question about the possibility of dynamical symmetry breaking.

Usually for the evaluation of the effective potential in gauged NJL
model one can use  Schwinger-Dyson equation \cite{15}. However this
equation, which presumbly gives the non-perturbative results is well
understood only in the situation of no external fields. In some very
special circumstances (for example, weak coupled massless QED in an
external weak magnetic field \cite{16}, or weak gravitational field
\cite{17} ) one can generalize SD equation taking  into account also background
fields. However, in general situations (arbitrary external gauge or
gravitational fields) that doesn't seem to be possible \cite{17}.

However, in the model under consideration one may use RG technique
because we have RG formulation of the gauged NJL-model. It has been
shown \cite{5} that in flat space the RG improved  effective potential
coincides with the one which was obtained via ladder SD equation
\cite{15}. Hence we expect that our results will be equivalent to
results one can obtain from ladder SD equation , which is not
formulated yet in background gauge fields.

The technique to study RG improved effective potential (effective
Lagrangian) is quite well-known in situation where background fields
are not presented \cite{18}. It is also known how to generalize it
(for the case of an external gravitational field, see \cite{19}). So
we will not give any details of this RG improvement technique (see
\cite{18,19}).

Using the fact that effective potential satisfies the RG equation, one
can solve this equation by the method of characteristics. Then
\begin{equation}\label{13} 
V(g,y,\lambda,e,m^2,\xi,\sigma,\mu)= 
V(\bar{g}(t), \bar{y}(t),\bar{\lambda}(t),\bar{e}(t),\bar{m}^2(t), 
\bar{\sigma}(t), \mu e^t)\;.
\end{equation} 
\noindent
Here, the effective coupling constants $\bar{g}(t),....,\bar{m}^2(t)$ are defined by
eqs. (\ref{10},\ref{11}) at scale $\mu e^t,\; \sigma(t)$ is written in
\cite{5}. The choice for RG parameter $t$ will be discussed below. As
it stands the relation (\ref{12}) is too general to be used in any real
calculation . In other words, the boundary condition to define
$V$ at $t=0$ should be added. The one-loop effective
potential (\ref{4},\ref{5}) is convenient to be used as such boundary
condition. 

As first case, we will consider the effective potential (\ref{4}) as
boundary condition. Then RG parameter $t$ is defined in the same way
as in \cite{5} (from the condition of the vanishing of the logarithmic
term). The final answer will be given by RG improved effective
potential written in \cite{5} (at finite cut-off or at cut-off tends
to infinity, see (\ref{6},\ref{13}) or (\ref{6},\ref{16}) of \cite{5}) (for correction of
misprint in \cite{5}, see \cite{10}) plus H-dependent term.  Note that
H dependent term appears in RG improved potential in the combination 
\[
\frac{1}{2}H^2-H^2\frac{a\bar{e}^2(t)}{24\pi^2}
\left(\ln\frac{\bar{M}^2_F(t)}{\mu^2e^{2t}}
\right)\;.
\]
Then,
\begin{eqnarray*}
{\frac{(4\pi)^2}{2a}}~{\frac{V}{\mu^4}} 
&=&     {\frac{1}{2}} 
        \left[{\frac{1}{g_{4R}(\mu)}}- 
                {\frac{1}{g_{4R}^{\ast}}}
\right]\frac{y^2_*\sigma^2(\mu)}{\mu^2}
+\frac{\alpha_c}{4\alpha} 
  \left(1+\frac{3\alpha}{2\alpha_c}\right) 
 ~\left[ \frac{y_*\sigma(\mu)}{\mu}\right]^{4/2-W}\nonumber\\ 
&{}& 
+\frac{1}{2}\frac{H^2}{\mu^4}\frac{(4\pi)^2}{2a}\;, 
\end{eqnarray*} 
\noindent
where $g^*_4 \equiv W,\; y_*$ denotes $y_\Lambda (t)$ (\ref{10}) at $\Lambda
\rightarrow \infty$. We wrote explicitly renormalized value of
effective potential. For $H=0$ it coincides with the result of refs.
\cite{5,15}. Hence, one can see that in such approach there
is no effect of magnetic field to chiral symmetry breaking, which may
occur in the case without external field.

Next, we analyse RG improved effective potential for the function (\ref{5})
chosen as boundary condition. With the condition of vanishing of
logarithmic terms in the effective potential (\ref{5}) we find $t$ as
follows:
\begin{equation}\label{14} 
e^t=\left(\frac{eH}{\mu^2}\right)^{1/2}~. 
\end{equation} 
Then RG improved effective potential (\ref{13}) at finite cut-off is 
\begin{eqnarray}\label{15}
&&{\frac{(4\pi)^2}{2a}}~{\frac{V}{\mu^4}} =     {\frac{x^2}{2}} 
        \left({\frac{\Lambda^2}{\mu^2}}\right)^W 
        \left[{\frac{1}{g_4(\Lambda)}}-{\frac{1}{W}}\right]~ 
\nonumber\\ 
&&+{\frac{x^4}{4}} 
  \left({\frac{eH}{\mu^2}}\right)^{\frac{1}{2}} ~\left[{\frac{3}{2}} 
  +     {\frac{\alpha_c}{\alpha}} 
  -  {\frac{\alpha_c}{\alpha}} 
                \left({\frac{\sqrt{eH}}{\Lambda}} 
                \right)^{\frac{2\alpha}{\alpha_c}}   
  \right]+\frac{1}{2}H^2\frac{(4\pi)^2}{2a\mu^4} \;,
\end{eqnarray} 
\noindent
where  $x=\frac{y_{\Lambda}(\mu)\sigma_{\Lambda}(\mu)}{\mu}$.

At the limit $\Lambda \rightarrow \infty$ we obtain
\begin{eqnarray}\label{16} 
{\frac{(4\pi)^2}{2a}}~{\frac{V}{\mu^4}} 
&=&     {\frac{x^2_{\ast}}{2}} 
        \left[{\frac{1}{g_{4R}(\mu)}}- 
                {\frac{1}{g_{4R}^{\ast}}}\right]  
+~{\frac{x^4_{\ast}}{4}} 
  \left({\frac{eH}{\mu^2}}\right)^{\frac{-\alpha}{\alpha_c}} 
 ~\left[ {\frac{3}{2}}+{\frac{\alpha_c}{\alpha}} \right]\nonumber\\ 
&{}& 
+\frac{(4\pi)^2H^2}{4a\mu^4}~.    
\end{eqnarray} 
The main qualitative result of this calculation is that, background
magnetic field in the above described aapproach doesn't change
no-external filed condition of chiral symmetry breaking:
\begin{equation}\label{17}
\frac{1}{g_{4R}(\mu)}-\frac{1}{g_{4R}^*} < 0\;.
\end{equation}
Of course, the above result is caused only by the fact that we treated
external magnetic field as expansion over powers of H. One can proceed
using the exact result (\ref{8}). Then there are no problems, in principle,
to calculate the effective potential explicitly. However the final
result will be too complicated, (it cannot be presented with
analytical expressions) and not very instructive.

\section{Discussion}

In summary, we found the effective potential for gauged NJL model in
external magnetic field. It is shown that structure of effective
potential is changed if compare with the case of no magnetic field.
However the weak magnetic field doesn't influence chiral symmetry
breaking condition. The technique developed in this work opens the
way to the study of gauged NJL model at external gauge fields without
using of Schwinger-Dyson ladder equation.

It is interesting to note that one can consider gauged NJL model in
external gravitational and magnetic fields (see \cite{6} for non
gauged case). The situation of physical interest maybe magnetically
charged black hole. Then,  $R=0$ but $R_{\mu \nu \alpha \beta}$ (or $R_{\mu\nu}
$) maybe not zero.

In applying again derivative expansion also over powers of $R_{\mu \nu
\alpha \beta}$ we will find the following additional term 
\[
\left(\beta + \beta_1 \ln  \frac{M^2_F}{\mu^2}\right) R^2_{\mu \nu
\alpha \beta}
\]
\noindent
in the potential (\ref{4}). Proceeding as above  we will again see that
external gravitational field  which describes the black hole  doesn't
change the condition of chiral symmetry breaking. Of course, strong
gravitational fields for gravitational fields of other configurations
(say $R \neq 0)$ change drastically the phase structure of gauged NJL
model \cite{10}.

Finally, one can immediately extend this paper to the study of chiral
symmetry breaking in the strong magnetic field (making use of
(\ref{3}) for big enough $H$) or in the external constant electric field
where particle creation occurs \cite{8}.

{\bf Acknowledgments} - The work by SDO has been supported by
Colciencias (Colombia) and FAPESP (S\~ao Paulo, Brasil), the work by
Gitman has been supported by CNPq (Brasil).

\end{document}